\documentclass[10pt,letterpaper]{article}
\usepackage{opex3}

\begin{document}



\title{Goos-H\"{a}nchen shifts of an electromagnetic wave
reflected from a chiral metamaterial slab}

\author{W. T. Dong$^1$,  Lei Gao$^{1,*}$, and C.W. Qiu$^{2,3}$}

\address{$^1$ Jiangsu Key Laboratory of Thin Films, School of Physical Science and
Technology, Soochow University, Suzhou 215006, China \\
$^2$ Department of Electrical and Computer Engineering, National
University of Singapore,  4 Engineering Drive 3, Singapore 117576,
Singapore\\
$^3$ Research Laboratory of Electronics, Massachusetts Institute of
Technology, 77 Mass. Avenue, Cambridge, MA 02139, USA.~~ {\sc Email:
cwq@mit.edu}}

\email{$^*$ leigao@suda.edu.cn} 



\begin{abstract}
Applying Artmann's formula to the TE-polarized incident waves, we
theoretically show that the Goos-H\"{a}nchen (GH) shifts near the
angle of the pseudo-Brewster dip of the reflection from a slab of
chiral metamaterial can be greatly enhanced. The GH shifts are
observed for both parallel and perpendicular components of the
reflected field. In addition, it is found that the GH shifts
depend not only on the slab thickness and the incident angle, but
also on the constitutive parameters of the chiral medium. In
particular, when the incident angle is close to the critical angle
of total reflection for LCP wave, significant enhancement of the
GH shifts can be obtained. Finally, the validity of the
stationary-phase analysis is demonstrated by numerical simulations
of a Gaussian-shaped beam.
\end{abstract}




\section{Introduction}
The Goos-H\"{a}nchen (GH) effect \cite{Goos1,Goos2}, which usually
refers to the lateral shift of the totally internal reflected from
the position predicted by geometrical optics, has been widely
analyzed both theoretically \cite{Artmann,Renard,Horowitz,Lai} and
experimentally \cite{Cowan,Bretenaker,Haibel}. This phenomenon has
already been extended to many fields such as acoustics, surface
optics, nonlinear optics, and quantum mechanics \cite{Emile,
Jost}. Furthermore, with the development of near-field scanning
optical microscopy and lithography \cite{Madrazo}, the importance
of the GH shift has been paid more and more attention. For
instance,  the GH shifts were found to be large positive or
negative for both reflected and transmitted beams in different
media or structures, such as dielectric slabs
\cite{Wang,Li,Yan,Huang}, metal surfaces \cite{Leung,Merano},
dielectric-chiral surface \cite{Hoppe,Depine}, absorptive media
\cite{Wild,Pfleghaar}, multilayered structures \cite{Tamir} and
photonic crystals \cite{Felbacq}. Recently, the GH shift
associated with left-handed material has been extensively studied
\cite{Berman, Kong, Lakhtakia, Qing, Chen, Xiang} owing to its
very unusual properties. On the other hand, since Pendry proposed
the Swiss roll structure to achieve a chiral medium with negative
refraction \cite{Pendry2}, negative chiral medium has been
receiving great interest
\cite{Tretyakov,Qiu_JOSAA,Qiu_PRB_chiral,Dong2}. In \cite{Dong1},
we have discussed the GH shift of electromagnetic waves incident
from an ordinary medium into negative chiral medium half space.
But, the thickness of chiral metamaterial is always finite, and
the properties of the GH shift and the reflection or transmission
will strongly depend on the thickness. Therefore, it is worth
discussing the effect of the thickness of the chiral metamaterial
on the GH shift and the contribution of the thickness and the
incident angle to the GH shift together. This is the main purpose
of the present paper.

In this paper, we study the GH shifts of the reflected waves from
a chiral metamaterial slab. We predict that the GH shift near the
angle of the pseudo-Brewster dip from such a slab can be large,
and both positive and negative lateral shifts are possible. It is
also shown that the GH shift depends on the thickness of the slab,
the incident angle of the wave and the constitutive parameters of
the chiral medium. Finally, numerical simulations for a
Gaussian-shaped beam have been performed to confirm the validity
of theoretical results. Here, only perpendicularly (TE) polarized
incident wave is discussed below, and the results for parallel
(TM) polarized incident wave can be easily obtained in the same
way.

\section{Formulation}

\begin{figure}[htb]
\centering\includegraphics[width=14cm]{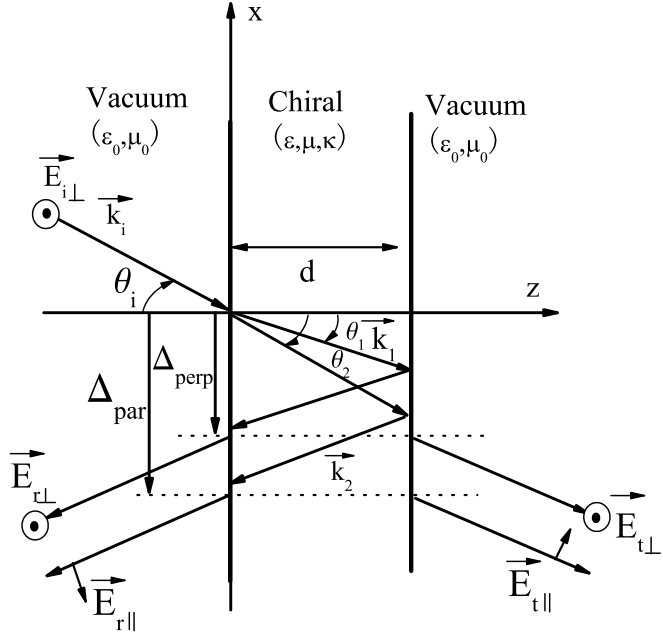} \caption{Schematic
diagram of a light beam propagating through the chiral slab placed
in free space.}
\end{figure}
The configuration for the chiral slab in a vacuum is shown in Fig.
1. We assume that a linearly polarized wave propagating in a
homogeneous isotropic dielectric medium is incident at an angle
$\theta_{i}$ upon the surface of a chiral slab with the thickness
$d$. The  constitutive relations of the chiral slab are defined by
\cite{Lindell}
\begin{eqnarray}
\mathbf{D} = \varepsilon \varepsilon _0 \mathbf{E} - i\kappa
\sqrt {\varepsilon _0 \mu _0 } \mathbf{H},  \\
\mathbf{B} = \mu \mu _0 \mathbf{H} + i\kappa \sqrt {\varepsilon _0
\mu _0 } \mathbf{E}, \label{eq:equation label}
\end{eqnarray}
where $\kappa$ is the chirality parameter (assumed to be positive
in this paper), $\varepsilon$ and $\mu$ are the relative
permittivity and permeability of the chiral medium, respectively
($\varepsilon _0 $ and $\mu _0 $ are the permittivity and
permeability in vacuum). For simplicity, a monochromatic
time-harmonic variation exp($i\omega t$) is assumed throughout
this paper, but omitted. It is seen that two interfaces of the
chiral slab are located at $z=0$ and $z=d$.

The incident electric and magnetic fields of an incident TE wave
can be written as
\begin{eqnarray}
\mathbf{E_i}  = E_0 \hat y \exp [ik_i ( - \cos \theta _i z + \sin \theta _i x)],~~~~~~~~~~~~~~~~~~~~~~~~~~~~~~\\
\mathbf{H_i}  = \sqrt {\frac{{\varepsilon _0 }}{{\mu _0 }}} E_0 (
- \cos \theta _i \hat x - \sin \theta _i \hat z) \exp [ik_i ( -
\cos \theta _i z + \sin \theta _i x)],
\end{eqnarray}
where $k_i=k_0 \equiv \omega \sqrt{\varepsilon _0 \mu _0}$ is the
wave number in the air. In a chiral material, an electric or
magnetic excitation will produce  both the electric and magnetic
polarizations simultaneously \cite{Silverman, Bassiri}. Therefore
the reflected wave must be a combination of the perpendicular and
parallel components in order to satisfy the boundary conditions.
According to the propagation direction of the reflected wave, the
electric and magnetic fields are expressed as
\begin{eqnarray}
\mathbf{E_r}  = E_0 [R_{11} \hat y + R_{21} \sin \theta _r \hat z
- R_{21} \cos \theta _r \hat x]\exp [ik_r (\cos \theta _r z + \sin
\theta _r x)],~~~~~~\\
\mathbf{H_r}  = \sqrt {\frac{{\varepsilon _0 }}{{\mu _0 }}} E_0
[R_{21} \hat y + R_{11} \cos \theta _r \hat x - R_{11} \sin \theta
_r \hat z]\exp [ik_r (\cos \theta _r z + \sin \theta _r x)] ,
\end{eqnarray}
where $\theta _r$ is the reflected angle, $k_r = k_i$ is the wave
number of the reflected wave, $R_{11}$ and $R_{21}$ are the
coefficients associated with perpendicular and parallel components
of the reflected wave, respectively.

Inside the chiral slab, there are two modes of propagation: a
right-circularly polarized (RCP) wave at phase velocity $\omega
/k_{1}$ and a left-circularly polarized (LCP) wave at phase
velocity $\omega /k_{2}$. The wave numbers $k_{1}$ and $ k_{2}$
have the form \cite{Lindell}
\begin{equation}
k_{1, 2}  = k_0 (\sqrt {\varepsilon \mu }  \pm \kappa ).
\end{equation}
Then, the refractive indices of the two eigen-waves in the chiral
medium are thus given as
\begin{equation}
n_{1, 2}  = \sqrt {\varepsilon  \mu }  \pm \kappa.
\end{equation}
Recent studies have shown that $\kappa>\sqrt {\varepsilon \mu }$
can occur at least at or near the resonant frequency of the
permittivity of a chiral medium (called chiral nihility
\cite{Tretyakov}), and then backward wave will occur to one of the
two circularly polarized eigen-waves, making negative refraction
in the chiral medium possible. When $\kappa>\sqrt {\varepsilon \mu
}$, the refraction index $n_{1}\equiv \sqrt {\varepsilon \mu } +
\kappa$ will still be positive, but the refraction index
$n_{2}\equiv \sqrt {\varepsilon \mu } - \kappa$ will become
negative. Correspondingly, negative refraction will arise for LCP
wave. Based on the discussion above, in the chiral slab, it is
assumed that there exist four total waves, two propagating toward
the interface $z=d$ and the other two propagating toward the
interface $z=0$ (see Fig. 1). The electric and magnetic fields of
these waves propagating inside the chiral medium toward the
interface $z=d$ are written as
\begin{eqnarray}
\mathbf{E_{c}^ + } = E_0 [A_1 \cos \theta _1 \hat x + A_1 \sin \theta _1 \hat z + iA_1 \hat y]\exp [ik_1 ( - \cos \theta _1 z + \sin \theta _1 x)] ~~~~~~\nonumber \\
  + E_0 [A_2 \cos \theta _2 \hat x + A_2 \sin \theta _2 \hat z - iA_2 \hat y]\exp [ik_2 ( - \cos \theta _2 z + \sin \theta _2 x)],~~~~~ \\
\mathbf{ H_{c}^ + }  = i\eta ^{ - 1} E_0 [A_1 \cos \theta _1 \hat x + A_1 \sin \theta _1 \hat z + iA_1 \hat y]\exp [ik_1 ( - \cos \theta _1 z + \sin \theta _1 x)] \nonumber \\
  - i\eta ^{ - 1} E_0 [A_2 \cos \theta _2 \hat x + A_2 \sin \theta _2 \hat z - iA_2 \hat y]\exp [ik_2 ( - \cos \theta _2 z + \sin \theta _2 x)].
\end{eqnarray}
The total electromagnetic fields of the other two waves
propagating inside the chiral medium toward the interface $z=0$
are expressed  as
\begin{eqnarray}
\mathbf{E_{c}^ - } = E_0 [ - B_1 \cos \theta _1 \hat x + B_1 \sin \theta _1 \hat z + iB_1 \hat y]\exp [ik_1 (\cos \theta _1 z + \sin \theta _1 x)] ~~~~~~\nonumber \\
  + E_0 [ - B_2 \cos \theta _2 \hat x + B_2 \sin \theta _2 \hat z - iB_2 \hat y]\exp [ik_2 (\cos \theta _2 z + \sin \theta _2 x)],~~~~~ \\
\mathbf{H_{c}^ - } = i\eta ^{ - 1} E_0 [ - B_1 \cos \theta _1 \hat x + B_1 \sin \theta _1 \hat z + iB_1 \hat y]\exp [ik_1 (\cos \theta _1 z + \sin \theta _1 x)] \nonumber \\
  - i\eta ^{ - 1} E_0 [ - B_2 \cos \theta _2 \hat x + B_2 \sin \theta _2 \hat z - iB_2 \hat y]\exp [ik_2 (\cos \theta _2 z + \sin \theta _2 x)],
\end{eqnarray}
where $\eta = \sqrt{\mu/\varepsilon}$ is the wave impedance of the
chiral medium, $A _{1,2}$ and $B _{1,2}$ are the transmitted
coefficients, and $\theta_{1,2}$ denote the refracted angles of the
two eigen-waves in the chiral slab, respectively.

Outside the slab ($z>d$), the total transmitted wave can be
expressed as 
\begin{eqnarray}
\mathbf{E_t} = E_0 [T_{11} \hat y + T_{21} \sin \theta _t \hat z +
T_{21} \cos \theta _t \hat x]\exp [ik_t ( - \cos \theta _t z +
\sin \theta _t x)], ~~~~~~\\
\mathbf{H_t} = \sqrt {\frac{{\varepsilon _0 }}{{\mu _0 }}} E_0
[T_{11} \hat y - T_{21} \sin \theta _t \hat z - T_{21} \cos \theta
_t \hat x]\exp [ik_t ( - \cos \theta _t z + \sin \theta _t x)],
\end{eqnarray}
where $k_t = k_i$, $\theta _t$ is the transmitted angle, $T_{11}$
and $T_{21}$ are coefficients associated with perpendicular and
parallel components of the transmitted wave in the air,
respectively.

The coefficients $R_{11}$, $R_{21}$, $A _{1,2}$, $B _{1,2}$,
$T_{11}$ and $T_{21}$ can be determined by matching the boundary
conditions at two interfaces $z=0$ and $z=d$, and the following
matrix can be obtained:
\begin{eqnarray}
\begin{array}{c}
  \left( {\begin{array}{*{20}c}
   {[\Psi]_{11}} & {[\Psi]_{12}}  \\
   {[\Psi]_{21}} & {[\Psi]_{22}}  \\
\end{array}} \right) \bullet \left( {\begin{array}{*{20}c}
   {R_{11} }  \\
   {R_{21} }  \\
   {T_{11} }  \\
   {T_{21} }  \\
\end{array}} \right) = \left( {\begin{array}{*{20}c}
   { - i\eta \cos \theta _i  + i\cos \theta _1 }  \\
   {i\eta \cos \theta _i  + i\cos \theta _1 }  \\
   {i\eta \cos \theta _i  - i\cos \theta _2 }  \\
   { - i\eta \cos \theta _i  - i\cos \theta _2 }  \\
\end{array}} \right). \\
 \end{array}
\end{eqnarray}
where
\begin{eqnarray}
\begin{array}{c}
   [\Psi]_{11}=\left( {\begin{array}{*{20}c}
     {- i(\eta \cos \theta _i  + \cos \theta _1 )} & { - \cos \theta _i  - \eta \cos \theta _1 }\\
     {i(\eta \cos \theta _i  - \cos \theta _1 )} & {\cos \theta _i  - \eta \cos \theta _1 }\\
     \end{array}} \right)\\
      \end{array}\\
\begin{array}{c}
      [\Psi]_{21} =\left( {\begin{array}{*{20}c}
     {i(\eta \cos \theta _i  + \cos \theta _2 )} & { - \cos \theta _i  - \eta \cos \theta _2 }\\
     {- i(\eta \cos \theta _i  - \cos \theta _2 )} & {\cos \theta _i  - \eta \cos \theta _2 }\\
     \end{array}} \right)\\
     \end{array}\\
\begin{array}{c}
[\Psi]_{12}  = \left( {\begin{array}{*{20}c}
     {i( - \eta \cos \theta _i  + \cos \theta _1 )e^{ - i(k_{iz}  - k_{1z} )d} } & {( - \cos \theta _i  + \eta \cos \theta _1 )e^{ - i(k_{iz}  - k_{1z})d}}\\
     {i(\eta \cos \theta _i  + \cos \theta _1 )e^{ - i(k_{iz}  + k_{1z} )d} } & {(\eta \cos \theta _1  + \cos \theta _i )e^{ - i(k_{iz}  + k_{1z})d}}\\
     \end{array}} \right)\\
     \end{array}\\
\begin{array}{c}
[\Psi]_{22} = \left( {\begin{array}{*{20}c}
     {i(\eta \cos \theta _i  - \cos \theta _2 )e^{ - i(k_{iz}  - k_{2z} )d} } & {( - \cos \theta _i  + \eta \cos \theta _2 )e^{ - i(k_{iz}  - k_{2z} )d}}\\
     { - i(\eta \cos \theta _i  + \cos \theta _2 )e^{ - i(k_{iz}  + k_{2z} )d} } & {(\eta \cos \theta _2  + \cos \theta _i )e^{ - i(k_{iz}  + k_{2z} )d} }\\
     \end{array}} \right).\\
\end{array}
\end{eqnarray}


The analytic solutions to the above matrix can be obtained after
some lengthy mathematic manipulations, but the final results are too
complicated to reproduce here. Now we would like to study the GH
shifts of the reflected beams from a slab of negative chiral medium.
It is well known that the lateral shift of the reflected beam has
the definition
\begin{equation}
\Delta  =  {d\Phi }/{{dk_x}},
\end{equation}
which was proposed by Artmann using the stationary-phase method
\cite{Artmann}. $\Phi$ is the phase difference between the
reflected and incident waves. For a chiral medium, the reflected
field contains both parallel and perpendicular components, these
field components can to first order be represented as two separate
reflected beams, each with its own magnitude and lateral shift.
Then the GH lateral shifts for perpendicular component ( $R_{11}$
in Eq. (15)) and parallel component ($R_{21}$ in Eq. (15)) of the
reflected wave can be calculated by the stationary-phase approach
as a function of the angle of incidence.

\section{Results and discussion}
\begin{figure}[htb]
\centering\includegraphics[width=14cm]{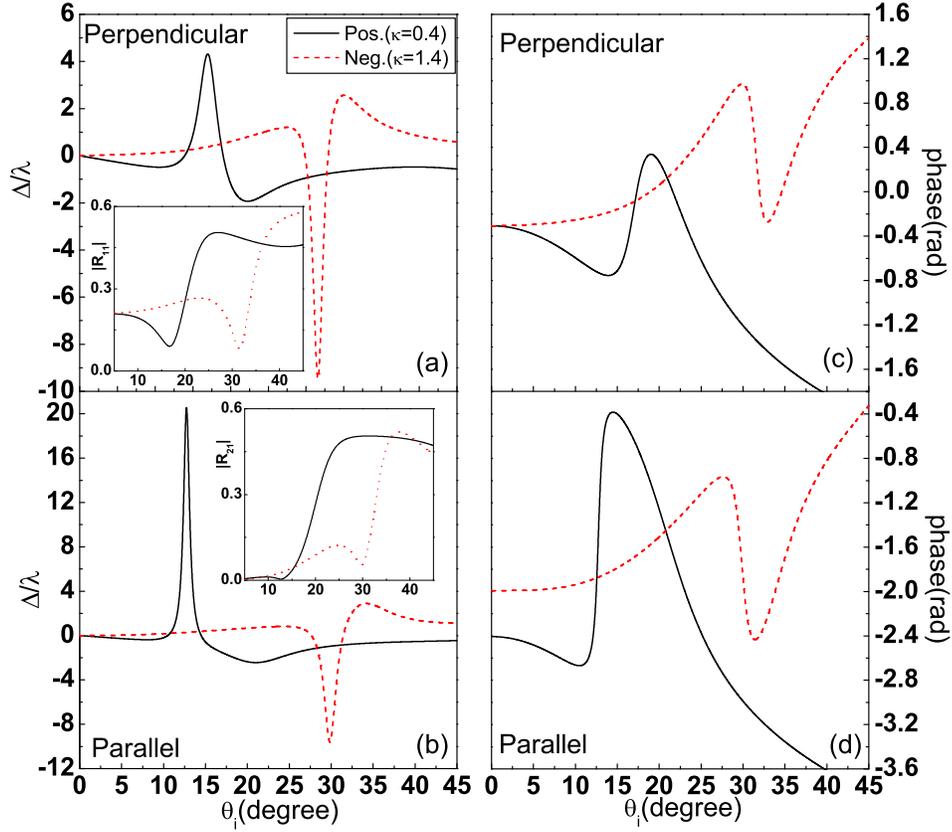} \caption{The
dependence of (a,b) the lateral shifts and (c,d) the phases of the
reflection coefficients for (a,c) perpendicular and (b,d) parallel
components on the incident angle in the presence of different
chiral slabs. The insets of (a) and (b) show the absolute values
of perpendicular and parallel reflection coefficients,
respectively. Solid line and dashed line correspond to positive
chiral slab and negative chiral slab, respectively.}
\end{figure}

In this section, numerical simulations associated with the GH
lateral shifts for a negative chiral slab will be presented and
compared with a conventional chiral slab.

In our calculation, we consider two types of chiral slabs: (1) a
positive chiral slab (e.g., $\varepsilon = 0.64$, $\mu = 1$, $\kappa
= 0.4$) whose refraction indices of RCP and LCP waves are both
positive, i.e, a conventional chiral slab; (2) a negative chiral
slab (e.g., $\kappa = 1.4$ and the other parameters unchanged) whose
refraction index of LCP wave is $n_2 = -0.6$. The working frequency
here is $\omega = 2\pi\times10$ GHz, and the thickness $d$ of the
chiral slab is $1.5 \lambda$. Figs. 2(a) and 2(b) show the
dependence of the GH shifts of the reflected beam upon the incident
angle for both perpendicular and parallel components from the slab
of Type-1 and Type-2. In Figs. 2(a) and 2(b), both the insets show,
respectively, the absolute values of perpendicular and parallel
reflection coefficients for two different types. It is easily found
that there is a dip in each reflection curve, at which $|R|$ reaches
the minimum (close to zero). The corresponding incident angle is
defined as the pseudo-Brewster angle. From Figs. 2(a) and 2(b), we
clearly see that the behaviors of the GH shifts for perpendicular
and parallel components are similar, and the shift will be greatly
enhanced near the angle of pseudo-Brewster dip. For the negative
chiral slab, the lateral shifts of both reflected components might
be large negative near the dip, and be small positive values under
other incident angles. This phenomenon can be easily explained in
terms of the change of phase. The phases of reflected parallel and
perpendicular components as a function of the incident angle are
plotted in Figs. 2(c) and 2(d) for negative and positive chiral
slabs. Near the angle of the dip, the phase of reflection
experiences a distinct sharp variation, which decreases quickly for
the chiral negative slab. As a result, Artmann's formula Eq. (16)
leads to a large negative GH shift and small positive lateral shift.
In contrast to chiral negative slab, for the positive chiral slab,
both components have positive lateral shifts near the angle of dip,
and then experience small negative shifts over other incident
angles. The dependence of GH shifts on incident angle for negative
chiral slab is opposite to that for positive chiral slab.
\begin{figure}[htb]
\centering\includegraphics[width=14cm]{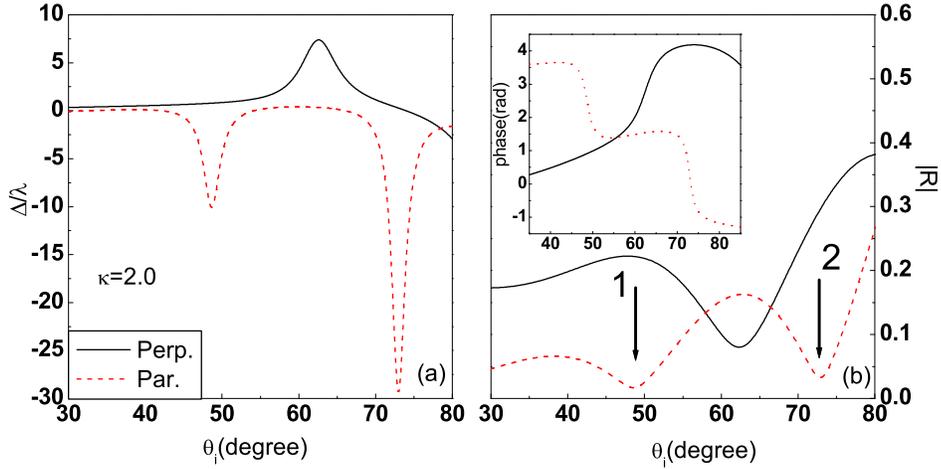} \caption{(a) The
dependence of the lateral shifts for reflected perpendicular and
parallel components on the incident angle for a typical negative
chiral slab. (b) The absolute values of perpendicular and parallel
reflection coefficients. The inset of (b) is the phase of both
reflection coefficients.}
\end{figure}

Figure 3 shows a typical dependence of the GH shifts of reflected
perpendicular and parallel components on the incident angle from a
negative chiral slab with a large chirality parameter. Assume that
the chiral slab has a chirality $\kappa=2.0$, while other parameters
remain the same as before. Figure 3(b) shows the absolute values of
the reflection coefficients for perpendicular and parallel
components as a function of the incident angle. It can be seen that
there is only one dip in the perpendicular reflection curve, at
which the reflection coefficient reaches a minimum magnitude and the
phase difference increases monotonically as a function of incident
angle (see the inset of Fig. 3(b)). Thus the GH shift of the
reflected perpendicular component has a positive peak near the angle
of the dip. In contrast, there are two dips in the parallel
reflection component, where the absolute values of the reflection
coefficient are very close to zero. Meanwhile, from the inset of
Fig. 3(b), the corresponding phase in the vicinity of these two dips
monotonically decreases quickly. Therefore, it indicates that the
shifts of the parallel component can be greatly enhanced to be large
negative near certain angles where the phase decreases. The shift
enhancement can be an order in magnitude greater than the
wavelength.
\begin{figure}[htb]
\centering\includegraphics[width=14cm]{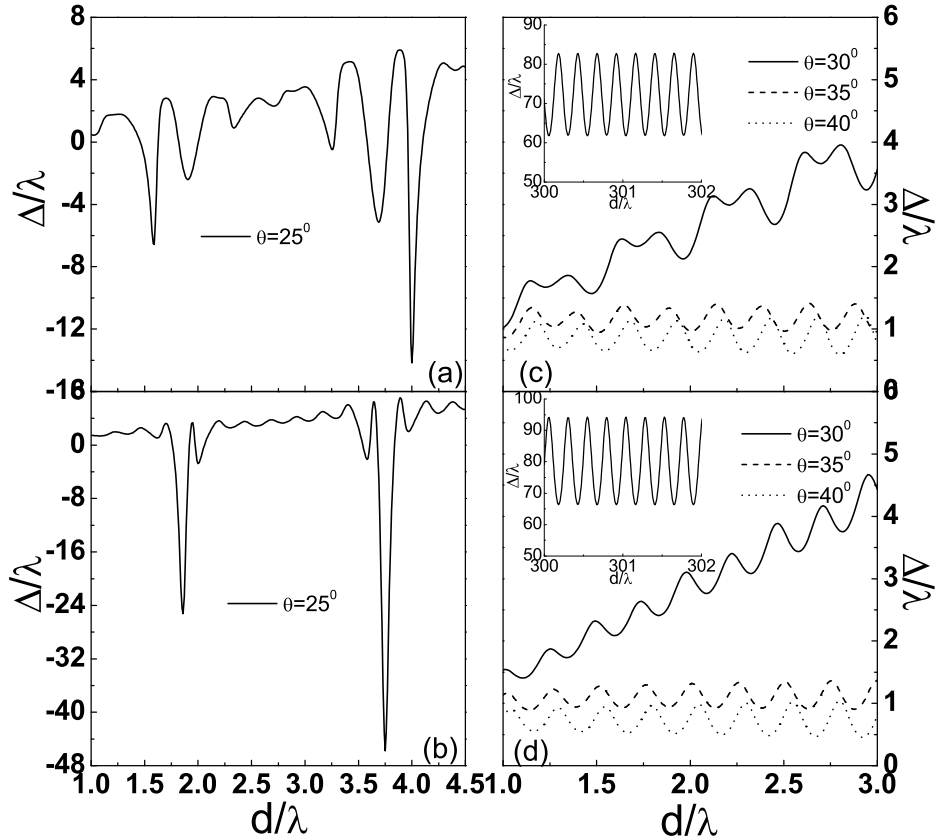} \caption{The
dependence of lateral shift on the thickness of the chiral
negative slab for different incident angles. (a) and (c) Reflected
perpendicular component; (b) and (d) Reflected parallel
component.}
\end{figure}

In what follows, we turn to discuss the lateral shift as a
function of the thickness of the slab under different incident
angles. First, we consider the GH shifts of the reflected beam
from a negative chiral slab. As an example, the dependence of
lateral shifts on the slab thickness at different incident angles
is presented in Fig. 4. The parameters are taken as follows:
$\varepsilon=0.64$, $\mu=1$, $\kappa=1.3$, implying that two
refraction indices corresponding to RCP and LCP waves are
$n_1=2.1$ and $n_2=-0.5$. It is obvious that there exists a
critical angle at $\theta_c=30^{\circ}$ for the LCP wave, but it
does not correspond to the true total reflection. Under such
critical situation, only LCP wave in the chiral slab becomes
evanescent wave when the incident angle exceeds this critical
angle, while the RCP wave will still propagate through the chiral
slab. Correspondingly, this incident angle is defined as
pseudo-Critical angle. However, the true total internal reflection
will never arise under these parameters. In such a case, when the
incident angle is smaller than the critical angle
($\theta_c=30^{\circ}$), it is shown in Fig. 4(a) that the GH
shift of the reflected perpendicular component could reach large
negative or positive values, but there never exists any periodic
fluctuation of the lateral shift with respect to the thickness.
Similar phenomenon can be found from the shift of the parallel
component (see Fig. 4(b)). Their large negative enhancement of the
lateral shifts correspond to the dips of the reflection
coefficients, which follows previous discussions. However, when
the incident angle is equal to the critical angle of the LCP wave,
as shown in Figs. 4(c) and 4(d), the lateral shifts of both
reflected components are always positive and increasing with
respect to the thickness of the slab. This can be explained by the
fluctuation of the reflection coefficient with the slab thickness.
However when the slab is much thicker, the strong fluctuation of
GH shift can be attained, as shown in the insets of Fig. 4. And
when the incident angle further increases above the
pseudo-Critical angle, the small fluctuation of the lateral shift
against the slab thickness is observed, as shown in Figs. 4(c) and
4(d).

\begin{figure}[htb]
\centering\includegraphics[width=14cm]{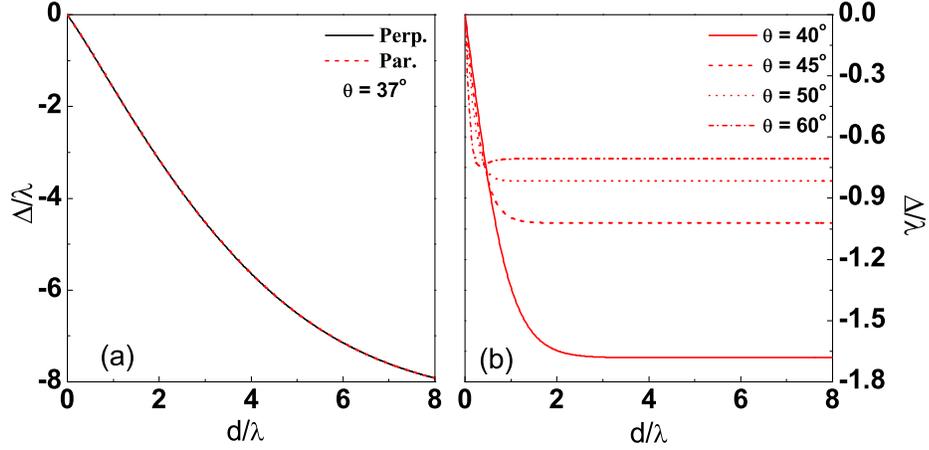} \caption{Dependence
of the GH shift on the thickness of an invisible medium slab for
different incident angles. $\varepsilon=0.8$, $\mu=0.8$, and
$\kappa=0.2$.}
\end{figure}

Apart from the above-mentioned the GH shift of the negative chiral
slab, we also consider the lateral shift of the other chiral
metamaterial slab. Here, we set the parameters of the chiral
medium as $\varepsilon=0.8$, $\mu=0.8$, and $\kappa=0.2$. In this
case, the wavenumber matching condition $k_1=k_0$ and the wave
impedance matching condition $\eta=\eta_0$ are satisfied
simultaneously, the RCP wave is transmitted straight through the
chiral medium without either reflection or refraction at any
incident angle. This medium is invisible for RCP wave
\cite{Tamayama}. While the LCP wave can be refracted and
reflected, or totally reflected. This unusual phenomenon can be
physically understood as a destructive interference of electric
and magnetic responses, due to the mixing through the chirality
parameter. For $k_2=0.6k_0$, Snell's equation for LCP wave is
expressed as $\sin\theta_i=0.6\sin\theta_2$, so the critical angle
for LCP wave is $\theta_c=\arcsin0.6\simeq37^{\circ}$. Therefore,
LCP wave is totally reflected with the incident angle greater than
$37^{\circ}$. This implies that we can divide the incident wave
into LCP and RCP waves. The property of GH shifts for this case is
shown in Fig. 5. Fig. 5(a) shows the perpendicular and the
parallel reflected GH shifts as a function of the slab thickness
when the incident angle is equal to the critical angle of the LCP
wave ($\theta_c=37^{\circ}$). Calculation under these conditions
shows that both reflected components have the same GH shift. This
is because the reflected wave only has LCP wave, and the RCP wave
contributes to the transmitted wave. Hence the perpendicular and
the parallel reflection coefficients have the same absolute value,
while their phases are different. From Fig. 5, it is
straightforward to see that the calculated lateral shifts for this
case are negative. Moreover, when the incident angle is close to
the critical angle of the LCP wave, the lateral shifts are large
and increase as the slab thickness increases. While the incident
angle is greater than $\theta_c$, we can see that, as the increase
of the slab thickness, the lateral shifts will increase quickly
and then gradually approach to an asymptotic negative value.
Therefore, the GH shift for the thick slab in the present case
depends on the incident angle more than the slab thickness.

\begin{figure}[htb]
\centering\includegraphics[width=10cm]{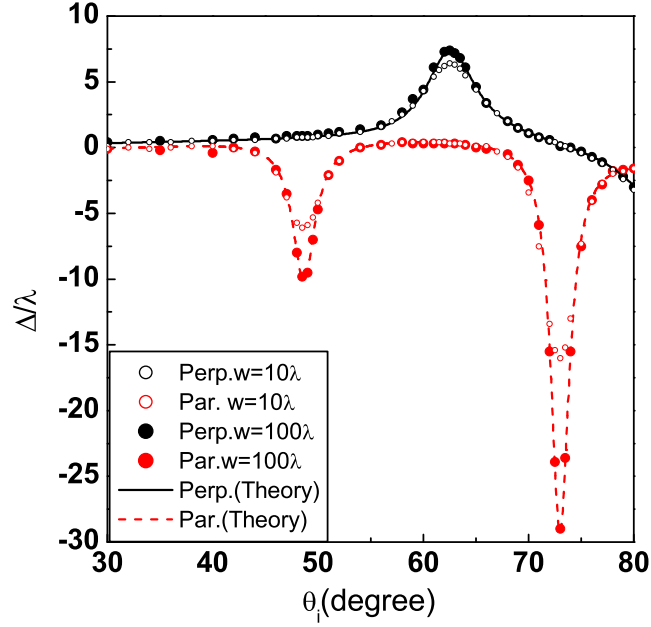} \caption{Dependence
of the GH shift on the incident angle. The theoretical result is
shown by the line; the numerical results are shown by solid
scatters (for $w_0=10\lambda$) and open scatters
($w_0=100\lambda$), all the other optical parameters are the same
as in Fig. 3(a).}
\end{figure}

\section{Goos-H\"{a}nchen shift of a Gaussian-shaped incident beam}
To show the validity of the above stationary-phase analysis,
numerical calculations are performed, which confirm our
theoretical results. In the numerical simulation, an incident
Gaussian-shaped beam is assumed,$ E_i (x,z = 0) = \exp ( - x^2
/2w_x^2  + ik_{x0} x)$, which has the Fourier integral of the
following form:
\begin{equation}
E_i (x,z = 0) = \int_{ - \infty }^\infty  {A(k_x )} \exp (ik_x
x)dk_x
\end{equation}
where $ w_x  = w_0 \sec \theta _i$, $w_0$is the beam width at the
waist, and the amplitude angular-spectrum distribution is
Gaussian, $A(k_x ) = \frac{{w_x }}{{\sqrt {2\pi } }}\exp [ -
\frac{{w_x^2 }}{2}(k_x  - k_{x0} )^2 ] $. So the electric fields
($E_ \bot ^r$ and $E_\parallel ^r$ ) of the reflected beam,
determined from the transformation of the incident beam, can be
written as
\begin{eqnarray}
E_ \bot ^r (x,z = 0) = \int_{ - \infty }^\infty  {R_{11} A(k_x )}
\exp (ik_x x)dk_x,~~~~~~~\\
E_\parallel ^r (x,z = 0) = \int_{ - \infty }^\infty  {R_{21} A(k_x
)} \exp (ik_x x)dk_x.~~~~~~~
\end{eqnarray}
The calculated beam shift can be obtained by finding the location
where $ \left| {E_ \bot ^r } \right|_{z = 0} $ or $\left|
{E_\parallel ^r} \right|_{z = 0} $is maximal \cite{Li}.

Calculations show that the stationary-phase approximation for the
GH shift is in good agreement with the numerical result. As an
example, Fig. 6 shows the numerical calculation results of curves
in Fig. 3(a). The incident beam width is chosen to be
$w_0=10\lambda$ and $w_0=100\lambda$. For comparison, both the
numerical and theoretical results are shown in Fig. 6. The peaks
of the numerical shifts for the perpendicular reflected field are
about $6.4\lambda $  for $w_0=10\lambda$ and $7.4\lambda $ for
$w_0=100\lambda$, and for the parallel reflected field are about
$-6.1\lambda$ (dip I), $-16\lambda$ (dip II) for $w_0=10\lambda$,
and $-9.8\lambda$ (dip I), $-29.0\lambda$ (dip II) for
$w_0=100\lambda$. The peaks of the theoretical shifts are about
$7.39\lambda$ for the perpendicular field and $-10.07\lambda$ (
$-29.32\lambda$) for dip I (dip II) of the parallel field. It is
noted that the discrepancy between theoretical and numerical
results is due to the distortion of the reflected beam, especially
when the waist of the incident beam is narrow \cite{Hsue, Li1}.
The further numerical simulation shows that the wider the incident
beam is, the less the discrepancy is, and the closer to the
stationary-phase result the numerical result is.

\section{Conclusion}
To summary, an investigation on the GH shifts of both reflected
parallel and perpendicular components  for  chiral metamaterial
slab has been done by using the stationary-phase approach. It
shows that the GH shift of the reflected perpendicular components
can be greatly enhanced to be large negative as well as positive
near the dip of the reflection curve, at which $|R|$ reaches a
minimum. Similar behavior will be found for the parallel
component. These results obtained from the negative chiral slab
are opposite to those from the conventional chiral slab. In
addition, at a given incident angle, the dependence of the GH
shift on the slab thickness for negative chiral slabs has also
been studied. It is shown that, when the incident angle is equal
to the critical angle of the LCP wave, the GH shifts of both
reflected components oscillate with respect to the thickness of
the slab, and its overall tendency is increasing along with the
small slab thickness, while for a much thicker slab, the
fluctuation of the lateral shift with the slab thickness becomes
more obvious. However, greatly enhanced GH shifts for a smaller
thickness slab can be obtained when the incident angle is smaller
than the critical angle. Meanwhile, we calculated the GH shift of
an invisible chiral medium for RCP wave. Finally, in order to
demonstrate the validity of the stationary-phase approach,
numerical simulations are made for a Gaussian-shaped beam.
Calculation results show that the wider the waist of the incident
beam is, the closer to the stationary-phase result it is.

\section*{Acknowledgments}
This work was supported by  the National Natural Science
Foundation of China under Grant No.~10674098, the National Basic
Research Program under Grant No.~2004CB719801, the Key Project in
Science and Technology Innovation Cultivation Program of Soochow
University, and the Natural Science of Jiangsu Province under
Grant No.~BK2007046.

\end{document}